\documentclass[aps,prb,twocolumn,superscriptaddress,showpacs]{revtex4-1}
\usepackage{graphicx}
\begin{document}
\title{Temperature and time scaling of the peak-effect vortex configuration in FeTe$_{0.7}$Se$_{0.3}$}
\author{Marco Bonura}\email{Marco.Bonura@unige.ch}
\affiliation{D\'{e}partement de Physique Appliqu\'{e}e - Universit\'{e} de Gen\`{e}ve, rue de l'\'{E}cole de M\'{e}decine 20, 1211 Gen\`{e}ve, Switzerland}
\author{Enrico Giannini}
\affiliation{D\'{e}partement de Physique de la Mati\`{e}re Condens\'{e}e, Universit\'{e} de Gen\`{e}ve, Quai Ernest-Ansermet 24, 1211 Gen\`{e}ve, Switzerland}
\author{Romain Viennois} \affiliation{D\'{e}partement de Physique de la Mati\`{e}re Condens\'{e}e, Universit\'{e} de Gen\`{e}ve, Quai Ernest-Ansermet 24, 1211 Gen\`{e}ve, Switzerland} \affiliation{Institut Charles Gerhardt Montpellier UMR 5253, Universit\'{e} Montpellier II and CNRS, Place Eug\`{e}ne Bataillon, 34095 Montpellier, France}
\author{Carmine Senatore}
\affiliation{D\'{e}partement de Physique Appliqu\'{e}e - Universit\'{e} de Gen\`{e}ve, rue de l'\'{E}cole de M\'{e}decine 20, 1211 Gen\`{e}ve, Switzerland}
\affiliation{D\'{e}partement de Physique de la Mati\`{e}re Condens\'{e}e, Universit\'{e} de Gen\`{e}ve, Quai Ernest-Ansermet 24, 1211 Gen\`{e}ve, Switzerland}
\newcommand{\Fe}{FeTe$_{0.7}$Se$_{0.3}$~}
\begin{abstract}
An extensive study of the magnetic properties of FeTe$_{0.7}$Se$_{0.3}$ crystals in the superconducting state is presented. We show that weak collective pinning, originating from spatial variations of the charge carrier mean free path ($\delta l$ pinning), rules in this superconductor. Our results are compatible with the nanoscale phase separation observed on this compound and indicate that in spite of the chemical inhomogeneity spatial fluctuations of the critical temperature are not important for pinning. A power law dependence of the magnetization vs time, generally interpreted as signature of single vortex creep regime, is observed in magnetic fields up to 8~T. For magnetic fields applied along the $c$ axis of the crystal the magnetization curves exhibit a clear peak effect whose position shifts when varying the temperature, following the same dependence as observed in YBa$_2$Cu$_3$O$_{7-\delta}$. The time and temperature dependence of the peak position has been investigated. We observe that the occurrence of the peak at a given magnetic field determines a specific vortex configuration that is independent on the temperature. This result indicates that the influence of the temperature on the vortex-vortex and vortex-defect interactions leading to the peak effect in FeTe$_{0.7}$Se$_{0.3}$ is negligible in the explored range of temperatures.

\end{abstract}
 \maketitle
%
\section{INTRODUCTION}
The study of the vortex properties in type-II superconductors is of extreme interest both for investigating the basic physics of the superconductivity and for evaluating the quality of the materials in view of practical applications. Many aspects of this subject are still topical, especially with regard to the high temperature superconductors (HTS) and the recently discovered iron-based superconductors. To date, five families of Fe-based superconductors have been discovered: \textit{RE}OFeAs, (''1111'', \textit{RE}=rare earth),\cite{Kamihara} \textit{A}Fe$_2$As$_2$ (''122'', \textit{A}=alkaline earth), \cite{Rotter} \textit{X}FeAs (''111'', \textit{X}=Li; Na),\cite{Wang,Parker} Fe(Se,Ch) (''11'', Ch=S, Te) \cite{Hsu,Yeh} and the most recently discovered ''21311'' family of Sr$_2$MO$_3$FePn (M=Sc, V, Cr and Pn=pnictogen).\cite{Ogino} Among these families, iron chalcogenides are considered of particular interest because of their simple crystal structure consisting of Fe ions tetrahedrally coordinated by Se and Te arranged in layers stacked along the c-axis, without any other interlayer cations, as occurs in the pnictides. For this reason, iron chalcogenides are generally considered an ideal candidate for understanding some open issues of high-temperature superconductivity.

One of the most intriguing phenomena observed in the study of the vortex properties in type-II superconductors is the so called peak (or fishtail) effect. A second peak in the magnetization curve $M$($H$) of low-$T_c$ superconductors has been observed as early as in the sixties\cite{De Sorbo} and a first theoretical explanation has been suggested by Pippard in 1969.\cite{Pippard} The peak has been found in proximity of $H_{c2}$ in Nb$_3$Sn, \cite{Lortz} CeRu$_2$, \cite{Banerjee} NbSe$_2$, \cite{Banerjee} V$_3$Si\cite{Isino} and MgB$_2$. \cite{Pissas, Angst}\\
The peak effect has been clearly observed also in the cuprate superconductors but its features vary from one compound to the other. In YBa$_2$Cu$_3$O$_{7-\delta}$ crystals, the peak is broad and temperature dependent; its origin has been attributed to the presence of inhomogeneities in the oxygen content\cite{Klein,Wen,Tobos} and associated to different mechanisms.\cite{Krusin-Elbaum, Civale,Abulafia,Kokkaliaris,Rosenstein}
In the more anisotropic Bi$_2$Sr$_2$Ca$_{n-1}$Cu$_n$O$_{2n+4+d}$ the peak in the magnetization loop has usually been associated to a 3D to 2D transition of the vortex structure. \cite{Yang1,Lee} However, significant differences in the second-peak features have been observed within this family: in the highly anisotropic Bi-2212 and Bi-2223 the peak is sharp and temperature independent \cite{Khaykovivh,Piriou1,Piriou2,Tamegai,Avraham} while in Bi-2201 crystals it exhibits a strong temperature dependence. \cite{Piriou3}
Regarding the iron-based superconductors, the peak effect has been observed in FeAs-1111, \cite{Moore,Senatore,Proz-NJP} in FeAs-122 \cite{Yang2,Prozorov,Proz-PhisicaC,Shen,Salem} and in FeTe$_{1-x}$Se$_x$,\cite{Das} even if a clear understanding of the physical origin of the phenomenon in this family has not been achieved, yet. However, since it has been shown by small angle neutron scattering,\cite{Eskildsen,Inosov,Eskildsen_2} Bitter decoration\cite{Eskildsen} and magnetic force microscopy\cite{Inosov} that an ordered vortex lattice is not present in these compounds, the origin of the peak effect due to a phase transition of the vortex lattice seems to be excluded.

Beyond the investigation of the magnetization curves, an useful approach for investigating the vortex dynamics is the study of the relaxation processes of the magnetization. The critical state of the vortex lattice, which determines the hysteresis of the magnetization in type-II superconductors, is a metastable state. It follows that vortices tend to hop out of their pinning-potential well in order to reach the configuration of absolute minimum energy. Such motion usually arises from thermal activation, but it can also arise from quantum tunneling (at low temperatures) or can be stimulated by external perturbations, such as microwave shaking of the vortex lattice. \cite{Mikitik} Magnetic relaxation processes have been observed in various low-temperature superconductors. \cite{Yeshurun} However, the subject has become of even greater interest after the discovery of high-$T_c$ superconductors, because of the higher operating temperatures and of the small activation energies related to the short coherence length and large anisotropy. The concept of thermally-induced hopping of the flux lines has been first treated by Anderson and Kim. \cite{Anderson,AndersonKim} In the framework of their model a logarithmic dependence of the magnetization ($M$) on the time ($t$) is expected. This behavior has been verified in various superconductors, both low $T_c$ and high $T_c$; however, in many experiments, deviations from the logarithmic dependence of $M$ have been observed, indicating a failure of the approximations that bring this expectation. \cite{Yeshurun,Zola} In the framework of the collective pinning theory a more complex expression for the dependence of the potential energy barrier height (the so-called \textquotedblleft interpolation formula\textquotedblright ) is proposed: $U(J)=U_0/ \mu [(J_c/J)^ \mu -1]$ where $\mu$ is a parameter varying  as a function of the vortex-vortex and vortex-pinning center interaction. \cite{Feigelman,Blatter} For example, in three dimensional systems $\mu= 1/7$ when the creep is dominated by the motion of individual flux lines, $\mu= 3/2$ in the case of collective creep of small bundles and $\mu= 7/9$ when the bundle size is much larger than the Larkin correlated volume~\cite{Blatter}. In 1991, Vinokur, Feigel'man and Geshkenbein proposed a theoretical model for the thermally activated flux creep, assuming a logarithmic dependence of the activation energy on the critical current, $U(J)=U_0 ln(J_c/J)$. \cite{Vinokur} This dependence of $U$ over $J$ is a good approximation for the creep activation barrier in the single vortex creep regime (limit $\mu \rightarrow 0$) and provides a proper fit of the experimental $U(J)$ dependence observed in different superconductors. \cite{Zeldov,Maley,McHenry,Zhang} The divergence of $U(J)$ when $J\rightarrow 0$ can be understood in the context of the collective pinning considering that in the vortex-glass state vortex motion is possible only in the presence of a current. \cite{Yeshurun} In the case of this logarithmic dependence of the activation energy on the critical current, a power-law time dependence of the magnetization is expected. \cite{Vinokur}

In this paper we study the magnetic properties of FeTe$_{0.7}$Se$_{0.3}$ crystals in the superconducting state by investigating the dependence of the magnetic moment ($m$) on the applied magnetic field ($H$) the temperature ($T$) and the relaxation time ($t$). In Sec.~\ref{EXP} we report details on the examined samples and on the experimental techniques used. The experimental results are shown and discussed in two different sections. In Sec.~\ref{VortPin}, dedicated to the vortex pinning properties, we analyze the $m(H)$ curves and deduce the pinning mechanism in the investigated crystals. The magnetic relaxation processes are studied in Sec.~\ref{VortDyn}. Conclusions are reported in Sec.~\ref{conclusions}.
\section{SAMPLE AND EXPERIMENTAL}\label{EXP}

The FeTe$_{0.7}$Se$_{0.3}$ crystals investigated were grown by a modified Bridgman-Stockbarger method, starting from a nominal Fe:(Te,Se) ratio of 0.9:1. Details of the procedure used for preparing the samples are reported in Ref.~\cite{Viennois}. The refined composition of the samples is Fe$_{1.013}$Te$_{0.68}$Se$_{0.32}$, indicating a very low iron excess.\cite{Viennois} It has been shown that reducing the Fe excess in Fe$_{1+x}$Te$_{1-y}$Se$_{y}$ favors the occurrence of superconductivity and weakens the antiferromagnetic order. \cite{Viennois,Liu} However, according to the phase diagram of the compound, \cite{Okamoto} a little excess of Fe is needed for stabilizing the structure. It is also worth noting that, due to Se-doping for Te, less Fe is allowed to occupy the additional site, since both the effects of reducing $x$ and increasing $y$ result in shrinking and re-shaping the FeTe$_4$ tetrahedra; so, the lower is the Se content, the more difficult is to limit the Fe excess.\cite{Viennois}

\begin{figure}
\centering \includegraphics[width=8 cm]{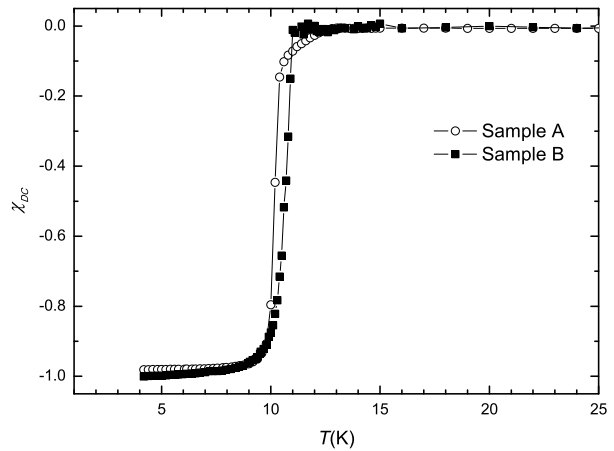} \caption{\label{chi}
Temperature dependence of the DC volume susceptibility (demagnetization corrected) for the investigated samples.}
\end{figure}

Two different crystals, weighting $\approx 50.5$ mg (Sample A) and $\approx 0.7$ mg (Sample B), have been investigated in the present study. The susceptibility curve $\chi (T)$ of the two samples has been deduced from a zero field cooled $m(T)$ measurement performed at 10 Oe by a Superconducting Quantum Interference Device (SQUID) with the field applied along the $c$ axis. Both curves have been corrected for the demagnetization effects according to the formula reported in Ref.\cite{Stoner}. The results are shown in Fig.\ref{chi}: the $\chi (T)$ curve of Sample A exhibits a superconducting transition with $T_c \equiv T(90\%) \approx 10.6$~K, $\Delta T_c \equiv T(90\%)-T(10\%) \approx 0.8$~K  whilst for Sample B it results $T_c \equiv T(90\%) \approx 10.9$~K, $\Delta T_c \equiv T(90\%)-T(10\%) \approx 1.0$~K. Both the samples exhibit bulk superconductivity with $\chi(4.25K)\approx -1$.

The samples have been characterized also by x-ray diffraction (XRD) both in a powder diffractometer (on manually ground crystals) and in a 4-circle diffractometer. The rocking curves indicate the presence of a distribution of the c-axis crystallites with a full width at half maximum (FWHM) of about $1.4^\circ$ for the Sample A and about $0.3^\circ$ for Sample B. Our results indicate that, whereas the sample B is of high crystalline quality, the sample A is more likely to be composed of few single-crystalline domains well aligned along the crystallographic c-axis.

The magnetization curves $m$($H$) of the \Fe samples were acquired by means of a SQUID magnetometer and a VSM (Vibrating Sample Magnetometer). The maximum applied magnetic field was 4.5~T for the SQUID and 8.8~T for the VSM. Measurements were performed in the range of temperatures from 4.25~K to $T_c$, with the magnetic field applied both parallel and perpendicularly to the $c$ axis. The VSM measurements were executed with different values of the magnetic field sweep rate ($SR$) namely 0.05~T/min, 1~T/min and 2~T/min. Furthermore, by means of the VSM, we studied the relaxation of magnetization over periods of time up to 6000 sec, for fixed values of temperature and magnetic field. In order to perform measurements in the trapped flux configuration, the relaxation measurements were performed with the following procedure: i) the sample is zero-field cooled; ii) the magnetic field is increased from zero to 8.8~T with $SR$= 2T/min; iii) the field is decreased from 8.8~T to 8~T with $SR$= 0.1~T/min; iv) the magnetic relaxation at 8~T is recorded for 6000 sec; iv) the field is decreased by 1 T with $SR$= 0.1~T/min; v) the magnetic relaxation is recorded for 6000 sec. Steps iv) and v) are repeated for registering the relaxation down to 0~T, in steps of 1~T.

\section{RESULTS AND DISCUSSION}\label{RESULTS}
\subsection{Vortex pinning properties}\label{VortPin}
\begin{figure}
\centering \includegraphics[width=8 cm]{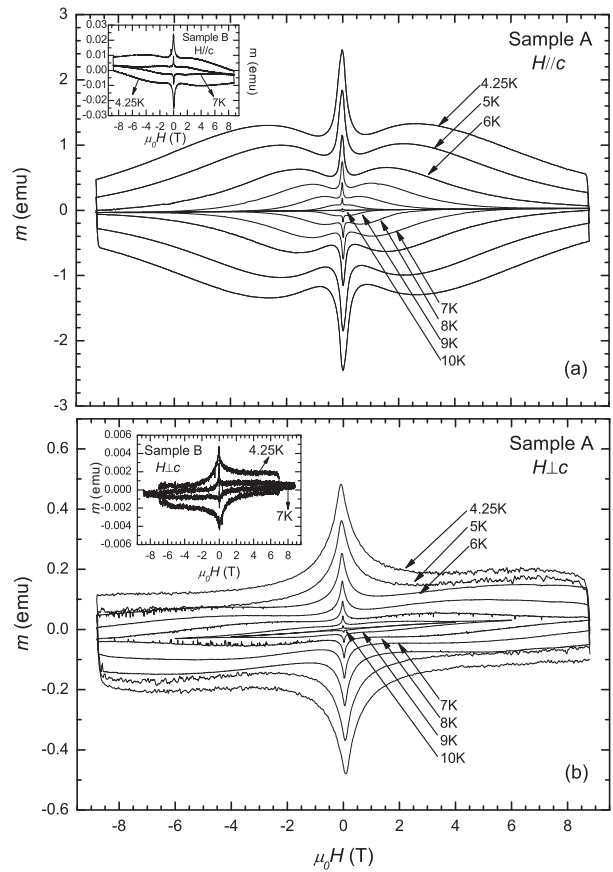} \caption{\label{m(H)}
Magnetic moment vs. external field curves obtained at different temperatures in Sample A for $H\parallel c$ (a) and $H\perp c$ (b). The insets show the results obtained in Sample B at $T=4.25$~K and $T=7$~K.}
\end{figure}

Fig.~\ref{m(H)} shows the magnetic-moment curves of the Sample A acquired with the VSM at different temperatures, for $H\parallel c$ (a) and $H \perp c$ (b), for a sweep rate of 2 T/min. Increasing the temperature, the pinning becomes weaker and consequently the width of the hysteresis loop decreases. A second peak in the $M$($H$) curve is clearly observed for $H\parallel c$, up to temperatures close to $T_c$.  A second peak is also present for $H \perp c$ but less pronounced and detectable only for temperatures higher than 5~K. In the inset of Fig.~\ref{m(H)} the results obtained for the Sample B at $T=4.2$~K and $T=7$~K, are reported for the two different orientations of the field. In this case, a second peak in the $m$($H$) curve is not observed for $H \perp c$, whereas it is clearly evident for $H\parallel c$, in analogy to what observed in YBa$_2$Cu$_3$O$_{7-\delta}$ crystals\cite{Klein} and in other Fe-based superconductors.\cite{Pramanik,Prozorov} Since Sample A presents a wider c-axis distribution of single-crystalline domains than Sample B, we argue that the weak peak effect revealed for $H \perp c$ in this sample is due to the slight misalignment of the crystallites forming the sample. We would also like to specify that the experimental $m(H)$ curves of Sample B for $H\parallel c$ are qualitatively similar with those of Sample A at comparable temperatures. This suggests that a misalignment of the layers forming the crystal is crucial only for magnetic fields applied perpendicularly to the $c$ axis of the sample. The $m(H)$ curves obtained for $H\perp c$ exhibit a paramagnetic-like background arising from the experimental setup. This background is not observed in the $m(H)$ curves measured for $H\perp c$ in the SQUID. In any case, the background does not alter the results obtained in the following, which rely only on the separation between the positive and negative branches of the magnetic-moment loop.

\begin{figure}
\centering \includegraphics[width=8 cm]{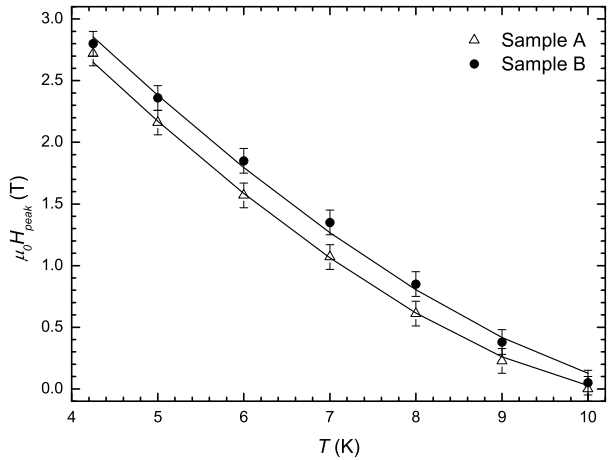} \caption{\label{h-peak(T)}
Temperature dependence of the magnetic field value at which the peak effect occurs, for the two measured samples. Continuous lines are the best fit curves obtained supposing the same $T$ dependence observed in YBa$_2$Cu$_3$O$_{7-\delta}$.}
\end{figure}

From the $m$($H$) curves measured for $H\parallel c$ at different $T$, we determined the temperature dependence of the magnetic field value corresponding to the second peak in magnetization ($H_{peak}$). The results for samples A and B are shown in Fig.~\ref{h-peak(T)}. It follows that for the FeTe$_{0.7}$Se$_{0.3}$ superconductor the position of the second-peak shifts towards lower fields monotonically on increasing the temperature, in analogy to what observed in the YBa$_2$Cu$_3$O$_{7-\delta}$ superconductor. \cite{Zhukov} For both samples investigated, the peak becomes undetectable at $T \approx 10$~K. The continuous lines in Fig.~\ref{h-peak(T)} are the best fit curves obtained supposing the same $T$-dependence of $H_{peak}$ as observed in YBa$_2$Cu$_3$O$_{7-\delta}$:\cite{Klein} $ H_{peak}=A\cdot (1-T/T^*)^{3/2}$, where $A$ is a constant and $T^*$ is the temperature at which the peak is undetectable. The best fit parameters (with the relative statistical errors) are: $A=5.9 \pm 0.1$~T, $T^*=10.3\pm 0.1$~K for Sample A and $A=6.0\pm 0.2$~T, $T^*=10.8\pm 0.2$~K for Sample B. This remarkable similarity between the results obtained in the YBa$_2$Cu$_3$O$_{7-\delta}$ and in the FeTe$_{0.7}$Se$_{0.3}$ superconductors, might suggest that the peak effect in the two systems has an analogous origin.

\begin{figure}
\centering \includegraphics[width=8 cm]{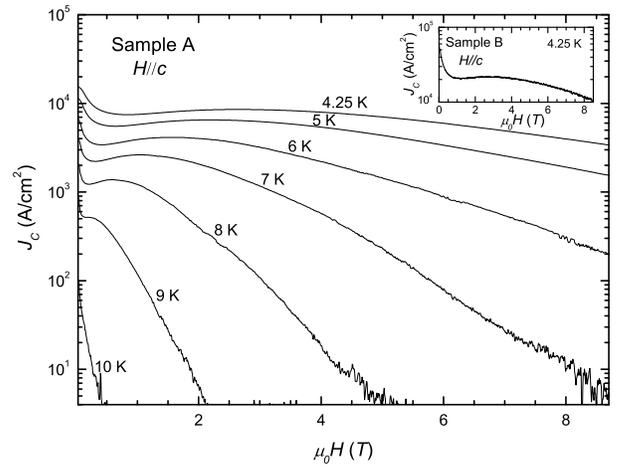} \caption{\label{Jc}
Field dependence of the critical current density at different temperatures and for $H \parallel c$, in Sample A. The inset shows the results obtained at $T=4.25$~K and for $H \parallel c$ for the Sample B.}
\end{figure}

The magnetic field dependence of the critical current density ($J_c$) has been extracted from the $m$($H$) curves, for different values of the temperature, using the Bean critical state formulae.\cite{Bean1,Bean2} For slab samples in a perpendicular magnetic field: $J_c(T,H)=3\Delta m(T,H) / w^2d(3l-w)$, where $\Delta m$($T,H$) is the separation between the two branches of the magnetic-moment loop, $l$ and $w$ are the length and the width of the sample ($l>w$) and $d$ is the thickness. The $J_c$($H$) curves obtained for the Sample A at different temperatures are presented in Fig.~\ref{Jc}; in the inset of the same figure the $J_c$($H$) curve at $T=4.25$~K of the Sample B is shown. For both samples investigated the $J_c$ values at 4.25~K are of the order of $10^4$~A/cm$^2$. This value is smaller than what measured in Fe$_{1+x}$Te$_{1-y}$Se$_{y}$ crystals with higher $y$ values than our samples.\cite{Taen,Liu2} However, the $J_c$ values become comparable if related to the same reduced temperature $T/T_c$.\cite{Liu2} The fact that the $J_c$ values obtained for Sample A are very similar to those obtained for the high-quality single crystal Sample B, where no weak links are present, indicates that weak links effects are not important for Sample A either.

The critical current density $J_c$ is always limited by the depairing current density $J_0=4 B_c/3\sqrt{6}\mu _0 \lambda$, where $B_c$ and $\lambda$ are the thermodynamic critical field and the GL penetration depth, respectively. \cite{Griessen} Important information can be deduced from the ratio $J_c/J_0$. For low-$T_c$ superconductors pinning is usually strong ($J_c/J_0\sim 10^{-2} - 10^{-1}$) resulting from the interaction of vortices with extended defects such as, e.g., precipitates or grain boundaries.\cite{Blatter} On the contrary, for the cuprate family pinning is usually weak ($J_c/J_0\sim 10^{-3} - 10^{-2}$) normally arising from point defects, e.g., oxygen  vacancies.\cite{Blatter} For the FeTe$_{0.7}$Se$_{0.3}$ we estimate $J_c/J_0\sim 10^{-3}$ indicating that also in the case of the Fe-based 11 family pinning is weak. The microscopic origin of pinning in Fe$_{1+x}$Te$_{1-y}$Se$_{y}$ will be discussed in the following.

As already mentioned, the presence of the peak effect in the magnetization curve has been widely documented in the literature, both in low and high $T_c$ superconductors. Its origin has been associated to different processes, depending on the particular system investigated. \cite{Pippard,Krusin-Elbaum,Rosenstein,Avraham} In order to shed light on the mechanisms that rule pinning in the Fe$_{1+x}$Te$_{1-y}$Se$_{y}$ superconductor, we have investigated the magnetic-field dependence of the pinning force density, $F_p$. It has been shown for a large variety of low-$T_c$ and high-$T_c$ superconductors that the curves of $F_p$ vs. $H$ obtained at different temperatures may be scaled into a unique curve if they are plotted as a function of the reduced field, $h=H/H_{irr}$.\cite{Fietz,Dew-Hughes,Koblischka1} The empirical formula that accounts for the scaling is:
\begin{equation}\label{Fp}
    F_p= C * h^ p (1-h)^ q\, ,
\end{equation}
where $C$ is a proportionality constant, $p$ and $q$ are two parameters whose values depend on the origin of the pinning mechanism. \cite{Dew-Hughes} The different contributions to flux-pinning are usually catalogued into two main categories: i) $\delta l$ (or normal) pinning, arising from spatial variations in the charge carrier mean free path ($l$); ii) $\delta T_c$ (or $\delta k$) pinning, associated with spatial variations of the Ginzburg parameter ($k$) due to fluctuations in the transition temperature $T_c$. \cite{Dew-Hughes, Blatter} In the last category fall also non-superconducting metallic particles whose dimension is smaller than the coherence length ($\xi$) since, in this case, the superconductivity is induced by the proximity effect. A classification is also made for pinning centers, as a function of the number of dimensions that are large with respect to the inter-vortex distance $d\approx (\phi _0 / B)^{0.5} $. Following the definition given by Dew-Hughes in Ref.\cite{Dew-Hughes}, in this paragraph we refer to point pins as regions whose dimensions in all directions are less than $d$, line pins have one dimension larger than $d$, grain- and twin- boundaries, which have two dimensions greater than $d$, act as surface pins, volume pins have all dimensions large with respect to $d$.\cite{Dew-Hughes}

In the framework of the Dew-Hughes model,\cite{Dew-Hughes} different values for $p$ and $q$ in Eq.~\ref{Fp} are expected, as a function of the specific pinning mechanism involved. Correspondingly, the theoretical $F_p$ vs. $h$ curves present a maximum at different $h$ values. In the case of $\delta l$ pinning, the maximum is expected at $h=0.33$ ($p=1$, $q=2$) for point pins and at $h= 0.2$ ($p=1/2$, $q=2$) for surface pins, such as grain boundaries (the same dependence has been predicted  for shear-breaking in the case of a set of planar pins \cite{Kramer}); no maximum is expected in the case of  $\delta l$ volume pinning, being $p=0$ and $q=2$. The maximum of the $F_p(h)$ curve is expected at higher $h$ values in the case of $\delta T_c$ pinning; in particular it occurs at $h=0.67$ ($p=2$, $q=1$) for point pins, at $h=0.6$ ($p=3/2$, $q=1$) for surface pins and at $h=0.5$ ($p=1$, $q=1$) for volume pins. Therefore, important information on the physical origin of the pinning mechanisms can be achieved by analyzing the scaled $F_p(h)$ curves.

\begin{figure}
\centering \includegraphics[width=8 cm]{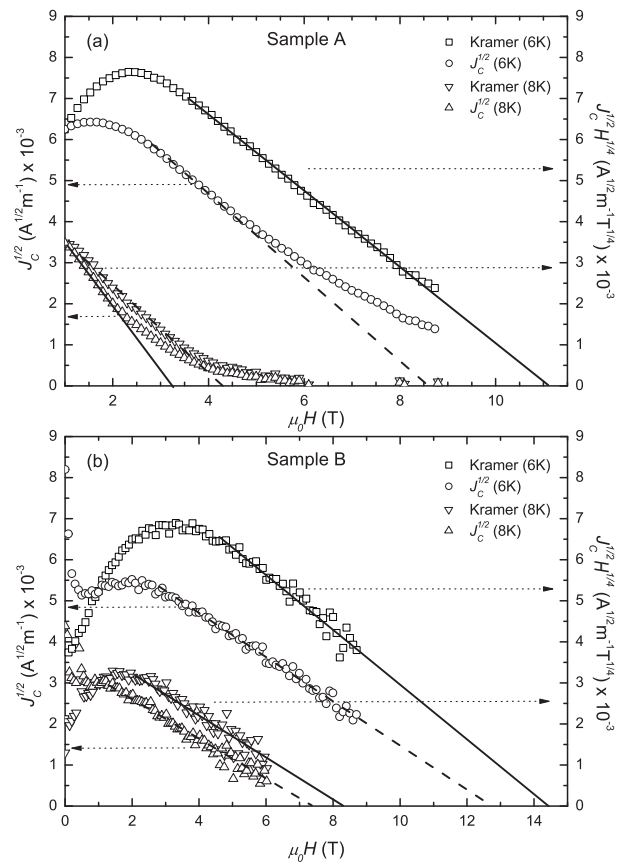} \caption{\label{KramerPlot}
Kramer plot (right axis) and $J_c^{1/2}$ vs. $H$ (left axis) curves obtained at $T=6$~K and $T=8$~K for sample A (a) and sample B (b), as described in the text.}
\end{figure}

\begin{figure}
\centering \includegraphics[width=8 cm]{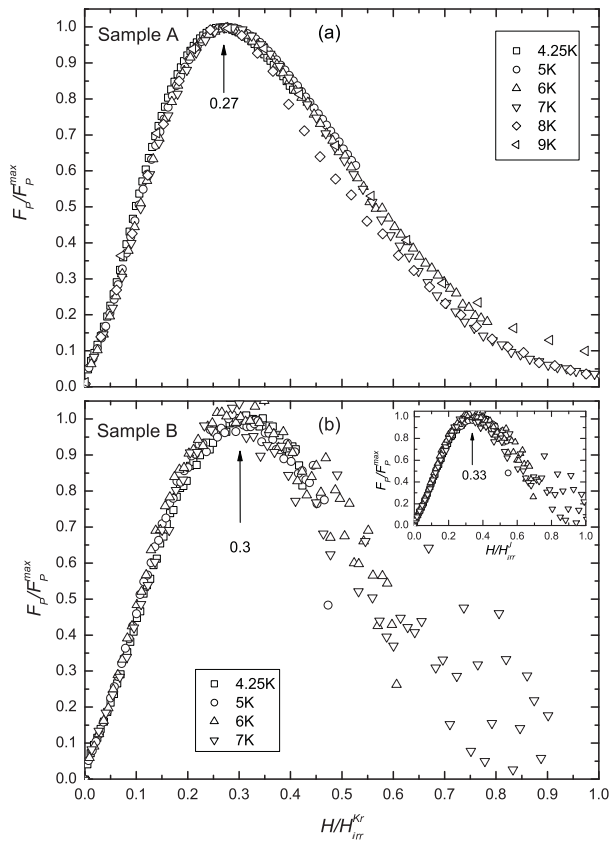} \caption{\label{FPinning}
Normalized pinning force density curves as a function of the reduced field, $H/H_{irr}^{Kr}$, for sample A (a) and B (b). $H_{irr}^{Kr}$ indicates that the irreversibilty field has been deduced by a linear extrapolation in the Kramer Plot. The inset in Fig.~\ref{FPinning} (b) shows the results obtained in sample B, scaled by using the irreversibility field, $H_{irr}^{J}$, deduced by the $J_c^{1/2}$ vs. $H$ curves }
\end{figure}

The comparison of the predicted pinning functions with the experiments requires an estimation of $H_{irr}$, defined as the $H$-value at which $J_c=0$. Starting from the experimental $J_c(H)$ curves, $H_{irr}$ is usually determined as the extrapolated zero value in the so-called Kramer plot,\cite{Kramer} where $(J_c^{1/2} \cdot H^{1/4})$ is plotted as a function of $H$. This procedure has been successfully used in the case of wires, polycrystalline samples and crystals.\cite{Flukiger,Ekin,Ekin2,Putti} However, if pinning is ruled only by defects whose dimensions are smaller than the intervortex distance, it is expected $J_c^{1/2} \propto (1-h)$.\cite{Dew-Hughes} In this case, $H_{irr}$ may be determined by doing a linear extrapolation down to zero of the $J_c^{1/2}$ vs. $H$ curve.\cite{Dew-Hughes,Montgomery,Shapira} In Fig.~\ref{KramerPlot} (a) and (b) we report the Kramer plots along with the $J_c^{1/2}$ vs. $H$ curves, obtained at two different temperatures, for Sample A and Sample B, respectively. We observe that the Kramer plot presents a wide linear behavior for both samples. On the other hand, the $J_c^{1/2}$ curve exhibits a linear behavior in a wide range of magnetic fields, only for Sample B. For this reason, we have extracted $H_{irr}$ from the Kramer plot for Sample A, whereas for Sample B we cannot discriminate a-priori which is the most correct procedure. However, a more conclusive result can be achieved if one considers simultaneously both the $J_c$ vs. $H$ dependence and the scaled $F_p (h)$ curves. If one hypothesizes that pinning is ruled by point pins and consequently extracts $H_{irr}$ from the $J_c^{1/2}$ vs. $H$ curve then the corresponding $F_p (h)$ curve should exhibit a maximum at $h \approx 0.33$.
Fig.~\ref{FPinning} (a) and (b) present the normalized pinning-force density for samples A and B at different temperatures as a function of the reduced field based on $H_{irr}$ values deduced from the Kramer plots, whereas the inset of Fig. 6 (b) is based on $H_{irr}$ values determined from a linear $J_c^{1/2}$ vs. $H$ dependence. For the sake of clearness, the superscript of $H_{irr}^{Kr}$ and $H_{irr}^{J}$ in the figure indicate respectively the $H_{irr}$ values extracted from the Kramer plot or hypothesizing $J_c^{1/2} \propto (1-h)$.

The pinning force curves of Sample A, obtained at different temperatures, scale well and present a maximum at $h\approx 0.27$ while, for Sample B, the maximum occurs at $h \approx 0.3$ if one considers $H_{irr} = H_{irr}^{Kr}$ or at $h \approx 0.33$ if $H_{irr} = H_{irr}^{J}$. These results indicate that, for the single crystal Sample B, the pinning mechanisms is ruled by defects whose dimensions are smaller that the intervortex distance in the investigated field range and as a consequence, the most appropriate procedure for determining $H_{irr}(T)$ is from the $J_c^{1/2}$ vs. $H$ curve. On the contrary, for Sample A the position of the maximum in the $F_p (h)$ curve suggests that a contribution to pinning coming from surface pinning cannot be excluded.
From the scaled pinning-force density curves we can also deduce information on the type of pinning centers. In particular, the fact that the maximum in the $F_p(h)$ curve occurs for values of the reduced field $\lesssim 0.33$ is an indication that the pinning centers in FeTe$_{0.7}$Se$_{0.3}$ are of $\delta l$ type. It is worth noting that this conclusion does not depend on the specific procedure used for determining H$_{Irr}$. In fact, in the case of pinning contribution coming from $\delta T_c$-type pins, the maximum is expected to occur at $h \gtrsim 0.5$, \cite{Koblischka} which is far away from what observed in our samples.

\begin{figure}
\centering \includegraphics[width=8 cm]{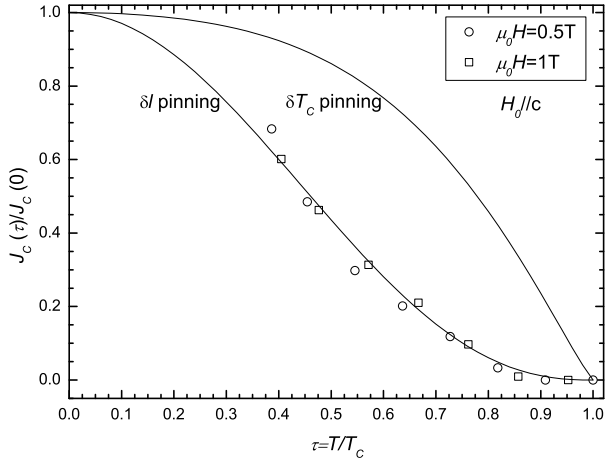} \caption{\label{GriessenGraph}
Normalized critical current density data, as a function of the reduced temperature ($\tau$), obtained at $\mu _0 H=0.5$~T and $\mu _0 H=1$~T for $H\parallel c$. The continuous lines are the theoretical curves expected in the case of $\delta l$ and $\delta T_c$ pinning.}
\end{figure}

The analysis of the scaled pinning force curves in the frame of the Dew-Hughes model revealed to give important information about pinning in low-$T_c$, cuprates as well as Fe-based superconductors.\cite{Koblischka1,Koblischka,Tobos,Yang2, Tarantini}. However since in our samples pinning is weak ($J_c/J_0\sim 10^{-3}$) and thermal fluctuations are large (Ginzburg number $\sim 10^{-3}$), more conclusive results about pinning origin in FeTe$_{0.7}$Se$_{0.3}$ are expected to be achieved by analyzing the experimental results in the framework of the collective pinning theory.\cite{Blatter} Following the theoretical approach proposed by Griessen \textit{et al}, \cite{Griessen} in the case of $\delta l$-type weak pinning in the single vortex regime it is expected that the critical-current density variation with respect to the reduced temperature ($\tau=T/T_c$) is described by the following expression:\cite {Griessen}

\begin{equation}\label{JC1}
    J_c(\tau)/J_c (0) = (1-\tau ^2) ^{5/2}(1+\tau ^2) ^{-1/2} \, ,
\end{equation}
while, for $\delta T_c$ pinning, it is:
\begin{equation}\label{JC2}
    J_c(\tau)/J_c (0) = (1 - \tau ^2) ^{7/6}(1 + \tau ^2) ^{5/6} \, .
\end{equation}
In Fig.~\ref{GriessenGraph}, we plot the normalized $J_c (\tau)$ data obtained at $\mu_0 H= 0.5$~T and $\mu_0 H= 1$~T for $H \parallel c$, along with the theoretical curves expected within the scenario of $\delta l$ and $\delta T_c$ pinning. The $J_c (\tau)$ values have been extracted from the $J_c$($H$) curves obtained at various temperatures. Analogous results have been obtained for sample B. A remarkably good agreement between the experimental results and the $\delta l$ pinning theoretical curve is obtained, confirming that pinning in the FeTe$_{0.7}$Se$_{0.3}$ samples originates form spatial variation of the mean free path.

The weak collective pinning theory considers that pinning originates from fluctuation in the density and force of defects whose dimension is smaller than the coherence length. In this case small values for the critical current density are expected ($J_c/J_0 \sim 10^{-3} - 10^{-2}$).\cite{Blatter} On the contrary, if pinning was mainly ruled by extended defects such as, e.g., precipitates, twinning or grain boundaries, or columnar defects such as dislocation lines, higher values for the $J_c/J_0$ ratio would be expected ($J_c/J_0 \sim 10^{-2} - 10^{-1}$).\cite{Blatter} A phase separation in Fe$_{1+x}$Te$_{1-y}$Se$_{y}$ single crystals with different $x$ and $y$ values has been experimentally observed by scanning tunneling microscopy and electron energy-loss spectroscopy. \cite{Hu,He} In particular, fluctuations of the local Te concentration (even by 20$\%$ from the average local composition) in nanometric regions have been observed in FeTe$_{0.7}$Se$_{0.3}$ crystals. \cite{Hu}  These fluctuations could be the origin of the observed weak pinning.

In spite of the inhomogeneous chemical distribution, our results indicate that local fluctuations of the critical temperature are not important for pinning. This interesting result is in agreement with tunneling spectroscopy experiments performed at $T=80$~K in optimal doped FeTe$_{0.55}$Se$_{0.45}$ crystals, indicating that the compound is chemically inhomogeneous but electronically homogeneous.\cite{He}

It is worth mentioning that a coexistence of superconducting and magnetic orders has been proposed for the Fe$_{1+x}$Te$_{1-y}$Se$_{y}$ system. In particular such coexistence has been inferred by Khasanov \textit{et al.}  based on muon-spin rotation experiments.\cite{Kasanov} On the other hand, Li \textit{et al.} have ascribed the apparent coexistence of superconductivity and magnetism in the Fe$_{1+x}$Te$_{1-y}$Se$_{y}$ system to a phase separation in the real space.\cite{Li} As already mentioned, a phase separation has also been observed in FeTe$_{0.55}$Se$_{0.45}$ crystals by STM.\cite{He}  Other authors have assumed a spin-glass state to exist between the long range antiferromagnetic order (at low Se content) and the bulk superconducting state, at high Se content.\cite{Katayama} The low-field DC susceptibility measurements of our samples, reported in Fig.\ref{chi}, do not show any anomaly or feature attributable to the coexistence of superconductivity and magnetic order. In a previous article we have deeply investigated the effect of the actual chemical composition and its effect on the crystal chemistry and on the magnetic and superconducting phase diagram.\cite{Viennois} We have proved the combined effect of Fe-excess ($x$) and Se-substitutions ($y$) on the transition from a superconducting to an antiferromagnetic state and highlighted the importance of controlling the real composition in a three dimensional phase diagram. By keeping under control a low Fe-excess, we have obtained bulk superconductivity at Se-content even lower than $0.3$.\cite{Viennois}  In 2010, Bendele \textit{et al.}\cite{Bendele2} has confirmed our study on the excess Fe and merged data from previous publications in a three dimensional phase diagram, still claiming a coexistence of superconductivity and magnetism. On the basis of our experimental results and the contradicting results reported in the literature, it is not possible to discriminate whether such coexistence occurs intrinsically at the atomic level in Fe$_{1+x}$Te$_{1-y}$Se$_{y}$ or a phase separation and local composition fluctuations are responsible for the magnetic behavior observed by some authors. Further studies on very clean and chemically homogeneous samples, carried out with different experimental techniques, are needed in order to clarify this point.

At the end of our analysis on the pinning properties we would like to remind that both the Dew-Hughes and the Griessen models have been developed in the single-vortex pinning regime, i.e. neglecting inter-vortex interactions. The validity of this approximation in our case is confirmed by the magnetic relaxation study reported in the next section.

\subsection{Vortex dynamics properties}\label{VortDyn}

\begin{figure}
\centering \includegraphics[width=8 cm]{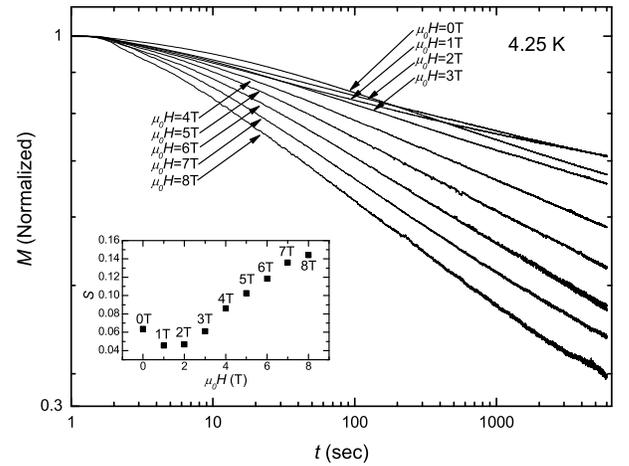} \caption{\label{FigRelax}
Plot of the magnetization vs. time (on a log-log scale) at $T=4.25$~K for different values of the magnetic field. The inset shows the field dependence of the relaxation rate $S=|$d~$ \log M/$d~$ \log t|$.}
\end{figure}

In this section we report a study of the dynamical properties of vortices in the FeTe$_{0.7}$Se$_{0.3}$ superconductor performed by magnetic relaxation measurements. The analysis has been limited to Sample A, since the measured magnetic moment of Sample B was too small for achieving a good resolution in a wide range of times and temperatures.
Fig.~\ref{FigRelax} shows the relaxation of the magnetization ($M=m/V$) normalized to its maximum value, obtained at $T=4.25$~K for $H\parallel c$, plotted in a log-log graph. At all the fields investigated and for times greater than about 50 sec, a linear dependence of $M$ has been observed, indicating a power-law dependence of $M$ vs. $t$. An analogous behavior has also been detected at higher temperatures and for $H \perp c$. It has been shown that deviations from the expected $M(t)$ curves at short times could be due to the magnetic field overshoot occurring when the external field ramp is stopped.\cite{Zola} This overshoot produces a shielded flux zone in proximity of the surface of the sample, which affects the initial relaxation process. The inset of Fig.~\ref{FigRelax} shows the $S=|$d~$ \log M/$d~$ \log t|$ values obtained at different $H$ values; large values of $S$ have already been observed in the Fe$_{1+x}$Te$_{1-y}$Se$_{y}$ system as well as in other Fe-based superconductors.\cite{Yadav,Prozorov,Yang-Sm} Since the $S$ value is related to the pinning potential energy barrier height, \cite{Blatter} the minimum observed in the $S$($H$) curve is associated to the presence of the peak effect in the $J_c$($B$) curves.

As already described in the introduction, in the framework of the Anderson and Kim theory, a linear dependence of $M$ on log($t$) is expected; this result stems from two basic assumption: i) the pinning potential energy barrier height decreases linearly with the current density: $U=U_0~(1-J/J_c)$; ii) $U_0/k_bT \gg 1$, which allows to hypothesize that the thermal-induced hopping rate is proportional to the Arrhenius factor $e^{-(U_0/k_bT)}$.
Supposing a linear dependence of $U$ on $J$ is only a first-order approximation whose validity has been demonstrated to fail many times. \cite{Yeshurun} As reported by Vinokur, Feigel'man and Geshkenbein, a power-law dependence of $M$ vs. $t$ is expected if a logarithmic dependence of the activation energy on the current density is supposed: $U(J)=U_0 \ln(J_c/J)$. \cite{Vinokur} In this case, the following expressions for the time dependence of $M$ are predicted:
\begin{equation}\label{VinExp}
    \ln(M)=cost-[x_f(t)/d](k_bT/U_0)\ln(t/\tau_0) \, , \textrm{for}~ t\ll t^*
\end{equation}
\begin{equation}\label{VinExp2}
    \ln(M)=cost - (k_bT/U_0) \ln(t/\tau_0) \, , \textrm{for}~ t\gg t^*
\end{equation}
where $d$ is the thickness of the sample, $x_f(t)$ is the position of the flux front and $t^*$ is the time at which the sample is fully penetrated. \cite{Vinokur,Zhang}

In order to verify that a logarithmic dependence of $U$($J$) is actually present for the FeTe$_{0.7}$Se$_{0.3}$ crystal investigated, we used the method proposed by Maley \textit{et al.},\cite{Maley} which allows the determination of the $U$($J$) curve from the experimental data obtained at different temperatures. In their paper, the authors show that choosing a proper value for a time-independent constant ($A$) it is possible to deduce the $J$ dependence of $U$ (or equivalently the $U$($M$) dependence) by plotting $U= -k_b T[\ln |$d$M/$d$t|-A]$ vs. $M-M_{eq}$, where M$_{eq}$ is the magnetization at the equilibrium. This procedure is valid under the assumption that the temperature dependence of $U$ is weak and that the principal effect of increasing the temperature is to produce monotonically decreasing initial values of $M$, which is usually valid for $T\lesssim T_c/2$. \cite{Maley}

\begin{figure}
\centering \includegraphics[width=8 cm]{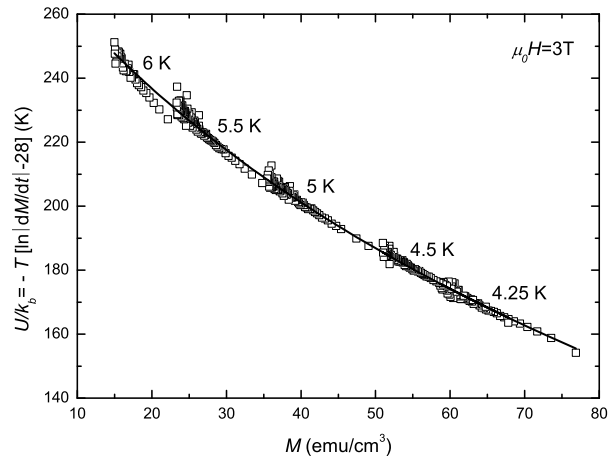} \caption{\label{FigMaley}
Magnetization dependence of the pinning potential energy barrier height calculated in the frame of the Maley model, scaling the data at different temperatures as described in the text.}
\end{figure}

Fig.~\ref{FigMaley} shows the curves obtained at $\mu _0 H = 3$~T, scaled considering A=28, along with the best fit curve deduced supposing the following dependence of $U$ on $M$: $U=U_0 \ln(a/(M-b))$; the best fit parameters (with the relative statistical errors) are: $U_0= 113 \pm 4$ K, $a= 5.0 \pm 0.2$ emu/cm$^3$ and $b= -0.39 \pm 0.03$ emu/cm$^3$. Similar values for $U_0$ have been obtained in other Fe-based superconductors, indicating that the weakness of pinning is a general characteristic of these compounds.\cite{Shen} The negative value of $b$ is due to the diamagnetic contribution of the sample holder. A very good scaling is obtained for $T\lesssim 5.5$ K, while at $T=6$ K a discrepancy between the experimental data and the theoretical curve starts to be present. As already mentioned in the introduction, the logarithmic dependence of $U$ on $J$ is a good approximation for the creep activation barrier in the single vortex creep regime. As a consequence, our results indicate that in the FeTe$_{0.7}$Se$_{0.3}$ superconductor the vortex motion develops in the single-vortex pinning limit even in magnetic fields up to 8 T. This is in general unexpected. At high fields, the intervortex distance becomes small compared to the magnetic field penetration depth and thus one would expect that vortex-vortex interactions become important and pinning involves vortex bundles. However, single vortex pinning up to 9 T has already been observed by Inosov \textit{et al.} in BaFe$_{2-x}$Co$_x$As$_2$ single crystals by magnetization measurements, small-angle neutron scattering, and magnetic force microscopy.\cite{Inosov} Our results suggests that also in the Fe$_{1+x}$Te$_{1-y}$Se$_{y}$ system the vortex-defect interaction dominates up to high fields. This experimental evidence further confirms the validity of our study on the pinning properties as reported in Sec.~\ref{VortPin}, carried out in the framework of the Dew-Hughes model where flux-lattice elasticity effects are neglected.

\begin{figure}
\centering \includegraphics[width=8 cm]{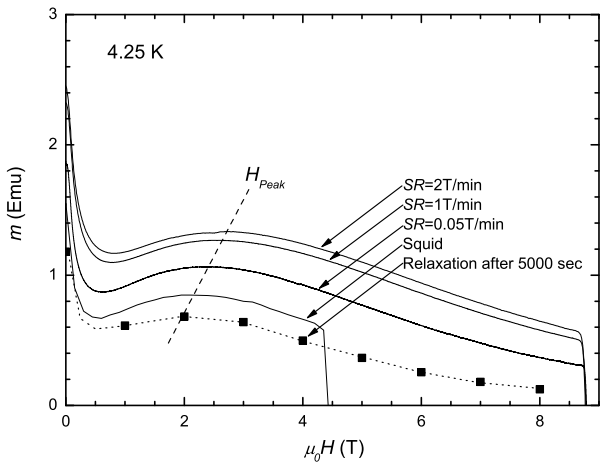} \caption{\label{FigPeakPosition}
Time evolution of the second peak position investigated by magnetization measurements performed by the VSM at different sweep rates ($SR$), by the SQUID as well as by magnetic relaxation measurements.}
\end{figure}

In Sec~\ref{VortPin} we have shown that the $m(H)$ curves obtained for $H\parallel c$ are characterized by a second peak whose position changes with the temperature. In order to investigate the magnetic relaxation effects on the peak effect, we measured the field dependence of the magnetic moment, by means of the VSM, for different sweep rates ($SR$) of the magnetic field. It has been demonstrated that the dependence of the hysteresis amplitude on the sweep rate contains basically the same information of the time dependence of the magnetization during relaxation, in particular, the higher is the sweep rate, the shorter is the effective ``observing time'' in the $m$($t$) curve.\cite{Schnack,Jirsa} The results are shown in Fig.~\ref{FigPeakPosition}, along with the curve acquired by the SQUID and the values (full points) extracted from the $m(t)$ curves, obtained at different fields, considering $t= 5000$ sec. Regarding the measurement performed with the SQUID, one can assume $SR=0$ since any variation of $H$ during the acquisition can be neglected. Furthermore, since the acquisition time is of some seconds, one has also to consider that the measured $m$ value has already relaxed. A remarkable variation of the $m(H)$ curves associated to different characteristic times is present, as a consequence of the relaxation in time of the magnetic moment. Furthermore, it is very interesting to note that the position of the second-peak moves towards low $H$ values during the relaxation, which indicates that the magnetic field value at which the second peak in the $J_c(H)$ curve is obtained relaxes in time.

\begin{figure}
\centering \includegraphics[width=8 cm]{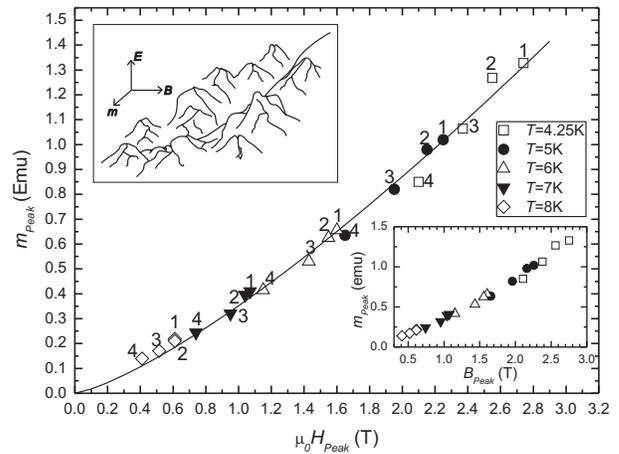} \caption{\label{Hpeak-Mpeak}
Pairs of values ($H_{peak}$,$m_{peak}$) that identify the second peak position in the $m(H)$ curve at different temperatures and characteristic times. Labels 1, 2, 3, indicate the measurements performed by the VSM with $SR$= 2T/min, 1T/min and 0.05 T/min, respectively; label 4 indicates the measurements performed by the SQUID. The continuous line is the best fit curve obtained as described in the text. The inset at the bottom right shows the couples of values ($B_{peak}$,$m_{peak}$) corresponding to the peak effect.  The inset at the top left is an imaginary drawing of the vortex energy landscape, the line indicating the relaxation path of the vortex configuration associated to the peak effect.}
\end{figure}

In order to shed light on the mechanism that determines the temperature and time dependence of the peak position, $H_{peak}$, we collected the couples of values ($H_{peak}$, $m_{peak}$) which identify its location in the $m$($H$) curves obtained at different temperatures by the SQUID and by the VSM and for different sweep rates, the results being shown in Fig.~\ref{Hpeak-Mpeak}. The inset at the bottom right of the same figure presents the calculated induction-field values ($B_{peak}$) corresponding to $m_{peak}$. It is very interesting to note that the pairs of values ($H_{peak}$, $m_{peak}$) - or equivalently ($B_{peak}$, $m_{peak}$) - obtained at different temperatures or sweep rates lie on the same curve and that, in some cases, the same pair of values ($H_{peak}$, $m_{peak}$) has been obtained at two different temperatures but for a different ``observing time'' during relaxation. The continuous line shown in the figure is the best fit curve obtained by assuming $m_{peak} = a \cdot \mu_0 H_{peak}^b$. The best fit parameters, along with the corresponding statistical errors, are: $a= 0.35\pm 0.01$~emu/T; $b=1.31\pm 0.04$.

Since the induction field $B$ is proportional to the number of vortices present in the SC and the magnetic moment identifies the critical current (and as a consequence the flux profile inside the sample), the pair of values ($B_{peak}$, $m_{peak}$) identify a specific configuration of the vortex structure. The fact that all the experimental data shown in Fig.~\ref{Hpeak-Mpeak} fall in a unique curve suggests that there is a unique vortex configuration which determines the occurrence of the peak effect at a given $H_{peak}$ value, whatever the temperature is in the range 4.2~K $\div$ 8~K, corresponding to 0.4 $T_c \div$ 0.75 $T_c$. In particular, the same ($B_{peak}$,$m_{peak}$) pair can be found at a temperature $T_1$ and observing time $t_1$ or equivalently at a temperature $T_2<T_1$ at a time $t_2>t_1$. If one defines the vortex energy landscape as the mapping of all the possible configurations of the vortices in the sample, determined by $B$ and $m$, and the corresponding energy level ($E$), the observed behavior could be justified by supposing that there is a unique path in the vortex energy landscape that describes the relaxation of the vortex configuration associated to the peak effect and that the temperature does not particularly affect the energy landscape in the $T$ range explored, at least in proximity of the path. The only effect of increasing the temperature is to allow the relaxation to start from a point in the path closer to the final equilibrium state. This result also indicates that the mechanism behind the second peak is related to a thermally driven process in the examined range of temperatures and fields. We are extending the present analysis of the peak effect properties to other superconductors, such as YBa$_2$Cu$_3$O$_{7-\delta}$, and the results will be discussed in a forthcoming paper.

\section{CONCLUSIONS}\label{conclusions}

We have investigated the magnetic properties of FeTe$_{0.7}$Se$_{0.3}$ crystals in the superconducting state by magnetization and magnetic relaxation measurements. We have shown that pinning in FeTe$_{0.7}$Se$_{0.3}$ originates form spatial variation of the mean free path and is most likely related to the nanoscale chemical phase separation observed in Fe$_{1+x}$Te$_{1-y}$Se$_{y}$ on a scale of $\sim 10$~nm.\cite{Hu,He} Very interestingly, our results confirm that even if chemically inhomogenous Fe$_{1+x}$Te$_{1-y}$Se$_{y}$ is electronically homogeneous since pinning is not ruled by spatial fluctuations of the critical temperature. From magnetic relaxation measurements we have obtained indications that vortex motion develops in the single vortex limit even in magnetic field as high as 8 T, in agreement to what observed in the 122 Fe-based compounds.\cite{Inosov}On applying the magnetic field along the $c$ axis, a clear peak effect in the $m(H)$ curves has been observed, up to temperatures near $T_c$. The second-peak position varies with the temperature, following the same dependence as observed in YBa$_2$Cu$_3$O$_{7-\delta}$.\cite{Klein} The relaxation of the vortex configuration that determines the peak effect has also been studied. We have found that the pairs of values ($H_{peak}$, $m_{peak}$) that identify the second-peak position in the $m(H)$ curves obtained at different temperatures and different relaxation times reconstruct a unique curve. This suggests that the vortex configuration that determines the peak effect at a particular $H$ value does not depend on the temperature in the range 0.4 $T_c$ $\div$ 0.75 $T_c$. It follows that the temperature influence on the vortex-vortex and vortex-defect interactions in proximity of the peak effect is negligible.

\section{ACKNOWLEDGEMENTS}
We acknowledge Prof. Thierry Giamarchi and Dr. Christophe Berthod for useful discussions, Dr. Florin Buta for critical reading of the paper, Damien Zurmuehle and Alexandre Ferreira for technical support.

\end{document}